\documentclass[11pt]{article}
\usepackage{amsmath}
\usepackage{amssymb}
\usepackage{bbm}
\usepackage{epsfig}
\allowdisplaybreaks[4]
\begin{document}
\title{Charge and CP symmetry breaking in two Higgs doublet models}
\author{A. Barroso$^{1,}$~\footnote{barroso@cii.fc.ul.pt},
P.M. Ferreira$^{1,}$~\footnote{ferreira@cii.fc.ul.pt} and 
R. Santos$^{1,}$~\footnote{rsantos@cii.fc.ul.pt}\\ 
$^1$ Centro de F\'{\i}sica Te\'orica e Computacional, Faculdade de Ci\^encias,\\
Universidade de Lisboa, Av. Prof. Gama Pinto, 2, 1649-003 Lisboa, Portugal}
\date{July, 2005} 
\maketitle
\noindent
{\bf Abstract.} We show that, for the most generic model with two Higgs doublets
possessing a minimum that preserves the $U(1)_{em}$ symmetry, charge breaking 
(CB) cannot occur. If CB does not occur, the potential could have two different 
minima, and there is in principle no general argument to show which one is the 
deepest. The depth of the potential at a stationary point that breaks CB or CP, 
relative to the $U(1)_{em}$ preserving minimum, is proportional to the squared 
mass of the charged or pseudoscalar Higgs, respectively.
\vspace{1cm}

Recently~\cite{pot10} we have been able to demonstrate a remarkable result: in 
the two Higgs doublet models (2HDM)~\cite{lee, sher} without explicit CP 
breaking, if, at the tree-level, there is a minimum that preserves the 
$U(1)_{em}$ and CP symmetries, that minimum is the global one. Hence, the 
stability of this minimum is guaranteed and tunneling to a deeper one that 
breaks charge conservation or CP becomes impossible. In this letter we will 
expand our analysis to the case of the 2HDM with explicit CP breaking, thus 
generalizing the conclusions of~\cite{pot10} to all possible potentials with two
Higgs doublets. 

There are many different ways of writing the 2HDM tree-level potential, but for 
the purpose of this work the best basis to work is the one introduced 
in~\cite{vel}. There are two scalar Higgs doublets in the theory, $\Phi_1$ and 
$\Phi_2$, both having hypercharges $Y=1$, i.e.,
\begin{equation}
\Phi_1 = \begin{pmatrix} \varphi_1 + i \varphi_2 \\ \varphi_5 + i \varphi_7
\end{pmatrix} \;\; , \;\; \Phi_2 = \begin{pmatrix} \varphi_3 + i \varphi_4 \\ 
\varphi_6 + i \varphi_8 \end{pmatrix} \;\; .
\end{equation}
The numbering of the real scalar $\varphi$ fields is chosen for convenience in
writing the mass matrices for the scalar particles. Our basis is obtained by 
first writing down the four $SU(2)_W \times U(1)_Y$ invariants one can construct
from these two doublets, namely
\begin{align}
x_1\,\equiv\,|\Phi_1|^2 &= \varphi_1^2 + \varphi_2^2 + \varphi_5^2 + \varphi_7^2
\nonumber \\
x_2\,\equiv\,|\Phi_2|^2 &= \varphi_3^2 + \varphi_4^2 + \varphi_6^2 + \varphi_8^2
\nonumber \\
x_3\,\equiv\,Re(\Phi_1^\dagger\Phi_2) &= \varphi_1 \varphi_3 + \varphi_2 
\varphi_4 + \varphi_5\varphi_6 + \varphi_7\varphi_8 \nonumber \\
x_4 \,\equiv\,Im(\Phi_1^\dagger\Phi_2) &= \varphi_1 \varphi_4 -  \varphi_2 
\varphi_3 + \varphi_5 \varphi_8 - \varphi_6 \varphi_7 \;\;.
\label{eq:x}
\end{align}
Notice that under a particular CP transformation, for a specific choice of basis
for the fields ($\Phi_1 \rightarrow \Phi_1^*\;,\;\Phi_2 \rightarrow \Phi_2^*$) 
the invariants $x_1$, $x_2$ and $x_3$ remain the same but $x_4$ changes sign. It
is now a simple matter to write down the most general 2HDM model, i.e., 
\begin{align}
V \;\;=& \;\; a_1\, x_1\, + \,a_2\, x_2\, + \,a_3 x_3 \,+\, a_4 x_4 \,+\, b_{11}
\, x_1^2\, +\, b_{22}\, x_2^2\, +\, b_{33}\, x_3^2\, +\, b_{44}\, x_4^2\, +\, 
\nonumber \\
 & \;\; b_{12}\, x_1 x_2\, +\, b_{13}\, x_1 x_3\, + b_{14}\, 
x_1 x_4\, +\,b_{23}\, x_2 x_3 +\,b_{24}\, x_2 x_4 +\,b_{34}\, x_3 x_4\;\; ,
\label{eq:pot}
\end{align}
where the $a_i$ parameters have dimensions of mass squared and the $b_{ij}$ 
parameters are dimensionless. On the whole, $V$ depends on 14 real
parameters although, with a particular choice of basis, one can reduce this 
number to 11 independent parameters (see, for instance,~\cite{hab}). The 
terms linear in $x_4$ are those that break CP explicitly, and eliminating them 
we are left with the CP preserving potential with 10 independent parameters that
we have used in reference~\cite{pot10}. In fact, with an appropriate choice of 
basis, this number of independent parameters may be reduced to 9~\cite{hab}. For
convenience we introduce a new notation, namely a vector A and a square matrix 
B, given by
\begin{equation}
A\;=\;\begin{bmatrix} a_1 \\ a_2 \\a_3 \\ a_4 \end{bmatrix} \;\;\; , \;\;\; 
B\;=\;\begin{bmatrix} 2 b_{11} & b_{12} & b_{13} & b_{14} \\ b_{12} & 2 b_{22} 
& b_{23} & b_{24} \\ b_{13} & b_{23} & 2 b_{33} & b_{34} \\ b_{14} &  b_{24} & 
b_{34} & 2 b_{44} 
\end{bmatrix} \;\; .
\label{eq:ab}
\end{equation}
Defining the vector $X\, =\, (x_1\,,\,x_2\,,\, x_3\,,\,x_4)$, we can rewrite the
potential~\eqref{eq:pot} as
\begin{equation}
V \;=\; A^T\,X \;+\; \frac{1}{2}\,X^T \,B\,X \;\;\; .
\label{eq:vm}
\end{equation}
It is a well known fact that the 2HDM potential has only three types of possible
minima~\cite{lee,sher}. A charge breaking (CB) minimum where, for instance, the 
fields $\varphi_5$, $\varphi_6$ and $\varphi_3$ have vevs. This last vev breaks
the $U(1)_{em}$ symmetry and so gives a mass to the photon. In the second type 
of minima only neutral fields have vevs, and there are two different 
possibilities. In one case only two fields ($\varphi_5$ and $\varphi_6$, for 
instance) have vevs, whereas in the second case three fields have vacuum 
expectation values ($\varphi_5$, $\varphi_6$ and $\varphi_7$, for example). We
call the first case the $N_1$ minimum, and the second the $N_2$ minimum. When
the model is reduced to the explicit CP preserving potential, $N_1$ is the 
CP preserving minimum and $N_2$ is the one that spontaneously breaks CP. For 
the potential where CP is explicitly broken from the start, there is no physical
distinction between the $N_1$ and $N_2$ minima. 

We now proceed to demonstrate that if one of the normal minima exists, it is 
certainly deeper than the charge breaking one. The demonstration follows 
closely our previous work~\cite{pot10}, so we refer the reader to that article 
for some non-essential intermediate steps. Let $V^\prime$ be a vector with 
components $V^\prime_i = \partial V/ \partial x_i$, evaluated at the $N_1$ 
minimum. At $N_1$ the non-zero vevs are $\varphi_5 = v_1$ and 
$\varphi_6 = v_2$, so that $x_1 = v_1^2$, $x_2 = v_2^2$, $x_3 = v_1 v_2$ and 
$x_4 = 0$. We define the vector $X_{N_1}$ as the value of $X$ at this minimum, 
and it is a trivial matter to demonstrate that the value of the potential at the
minimum is given by
\begin{equation}
V_{N_1} \;\; = \;\; \frac{1}{2}\,A^T\,X_{N_1} \;\; = \;\; -\frac{1}{2}\,
X^T_{N_1}\, B\,X_{N_1} \;\;.
\label{eq:vmin}
\end{equation}
Further, we can write down the non-trivial stationarity conditions, which are
\begin{equation}
\begin{array}{rclcl}
\displaystyle{\frac{\partial V}{\partial v_1}}= 0 &\Leftrightarrow & V^\prime_1 
\displaystyle{\frac{\partial x_1}{\partial v_1}} \,+\, V^\prime_3 \displaystyle{
\frac{\partial x_3}{\partial v_1}} = 0 &\Leftrightarrow &  V^\prime_1  = 
\left( \displaystyle{-\frac{V^\prime_3}{2 v_1 v_2}} \right)v_2^2\vspace{0.2cm}\\
\displaystyle{\frac{\partial V}{\partial v_2}} = 0 &\Leftrightarrow &  
V^\prime_2 \displaystyle{\frac{\partial x_2}{\partial v_2}} \,+\,V^\prime_3 
\displaystyle{\frac{\partial x_3}{\partial v_2}} = 0 &\Leftrightarrow &  
V^\prime_2  = \left(\displaystyle{-\frac{V^\prime_3}{2 v_1 v_2}}\right) v_1^2 
\vspace{0.2cm}\\
\displaystyle{\frac{\partial V}{\partial \varphi_7}} = 0 &\Leftrightarrow &  
V^\prime_4 \displaystyle{\frac{\partial x_4}{\partial \varphi_7}} = 0 
&\Leftrightarrow &  V^\prime_4 = 0 \;\;\;\;\;\; .
\label{eq:min}
\end{array}
\end{equation}
From eq.~\eqref{eq:vmin} we see that $V^\prime \,=\, A\,+\,B\, X_{N_1}$ and from
the equations above we can obtain
\begin{equation}
V^\prime \; = \; \begin{bmatrix} V^\prime_1 \\ V^\prime_2 \\ V^\prime_3 \\ 
V^\prime_4 \end{bmatrix} \; = \; -\frac{V^\prime_3}{2 v_1 v_2}\, \begin{bmatrix}
v_2^2 \\ v_1^2 \\ - 2 v_1 v_2 \\ 0 \end{bmatrix} \;\; . 
\label{eq:vl}
\end{equation}
Written in this form we see, plainly, that $V^\prime_1$ and $V^\prime_2$ have 
the same sign. Now, in reference~\cite{pot10} we have obtained general 
expressions
for the mass matrices of the theory's scalar particles (equations (31)-(35) of 
that paper), which do not depend on the particular form of the potential (to put
it more bluntly, they do not depend on the potential having 10 or 14 
parameters). In particular, the charged scalars' mass matrix when there are no 
charged vevs is given by
\begin{equation}
[M^2_{H^\pm}] \;=\; \begin{bmatrix} 2 V^\prime_1 & 0 & V^\prime_3 & V^\prime_4 
\\ 
0 & 2 V^\prime_1 & - V^\prime_4 & V^\prime_3 \\ V^\prime_3 & - V^\prime_4 & 2
V^\prime_2 & 0 \\ V^\prime_4 & V^\prime_3 & 0 & 2 V^\prime_2 \end{bmatrix} \;\;,
\label{eq:mm}
\end{equation}
Using eqs.~\eqref{eq:min} one can see that the non-zero eigenvalue of this 
matrix is $M^2_{H^\pm}\, =\, V^\prime_1\,+\,V^\prime_2\, =\, -V^\prime_3 v^2/(2 
v_1 v_2)$, with $v^2\,=\,v_1^2\,+\, v_2^2$. If $N_1$ is a minimum then all of 
the squared scalar masses are positive and so this quantity is positive. Another
consequence of the minimisation conditions is that we obtain $X_{N_1}^T\,
V^\prime \,=\,0$. 

Regarding the CB stationary point, the fields that have non-zero vevs are now
$\varphi_5 = v^\prime_1$, $\varphi_6 = v^\prime_2$ and $\varphi_3 = \alpha$. We
define the vector $Y$ to be equal to the vector $X$ evaluated at this stationary
point, that is, $Y$ has components $Y = ({v^\prime_1}^2 \,,\,{v^\prime_2}^2\,+\,
\alpha^2 \,,\,v^\prime_1 v^\prime_2 \,,\, 0)$. The stationarity conditions now 
give
\begin{equation}
\begin{array}{rclcl}
\displaystyle{\frac{\partial V}{\partial v^\prime_1}}= 0 &\Leftrightarrow & 
V^\prime_1 \displaystyle{\frac{\partial x_1}{\partial v^\prime_1}} \,+\, 
V^\prime_3 \displaystyle{\frac{\partial x_3}{\partial v^\prime_1}} = 0 
&\Leftrightarrow &  V^\prime_1  = \left( \displaystyle{-\frac{V^\prime_3}{2 
v^\prime_1 v^\prime_2}} \right) {v^\prime_2}^2 \vspace{0.2cm} \\
\displaystyle{\frac{\partial V}{\partial v^\prime_2}} = 0 &\Leftrightarrow &
V^\prime_2 \displaystyle{\frac{\partial x_2}{\partial v^\prime_2}} \,+\,
V^\prime_3 \displaystyle{\frac{\partial x_3}{\partial v^\prime_2}} = 0 
&\Leftrightarrow & V^\prime_2  = \left(\displaystyle{-\frac{V^\prime_3}{2 
v^\prime_1 v^\prime_2}}\right) {v^\prime_1}^2 \vspace{0.2cm} \\
\displaystyle{\frac{\partial V}{\partial \alpha}} = 0 &\Leftrightarrow &
V^\prime_2 \displaystyle{\frac{\partial x_2}{\partial \alpha}} = 0 
&\Leftrightarrow & V^\prime_2 = 0 \vspace{0.2cm} \\
\displaystyle{\frac{\partial V}{\partial \varphi_7}} = 0 &\Leftrightarrow &
V^\prime_4 \displaystyle{\frac{\partial x_4}{\partial \varphi_7}} = 0
&\Leftrightarrow &  V^\prime_4 = 0 \;\;\;\;\;\; .
\end{array}
\end{equation}
We thus obtain, for the CB stationary point, $V^\prime_i = 0$, just like the 
case of the 10 parameter potential treated in ref.~\cite{pot10}. The equation 
that determines $Y$ is simply $A\; + \; B\,Y \;\; = \;\; 0$, which implies that,
even for this more complex potential, the CB stationary point, if it exists, is 
unique. We can therefore follow the steps of the demonstration done 
in~\cite{pot10}, and observe that the value of the potential at this charge 
breaking stationary point is given by
\begin{equation}
V_{CB} \;\; = \;\; \frac{1}{2}\,A^T\,Y \;\; = \;\; -\frac{1}{2}\,Y^T\, B\,Y 
\;\;.
\label{eq:vcb}
\end{equation}
Remembering that $X_{N_1}^T\,V^\prime \,=\,0$ we obtain, from $V^\prime \,=\, 
A\,+\,B\, X_{N_1}$ and $A\; + \; B\,Y \;\; = \;\; 0$, that 
\begin{equation}
X_{N_1}^T\,B\, Y \;\; = \;\; X_{N_1}^T\,B\, X_{N_1} \;\; = \;\; -\,2\,V_{N_1}
\;\; .
\label{eq:vn1}
\end{equation}
We can calculate the quantity $Y^T\,V^\prime$, which is easily seen to be given
by
\begin{equation}
Y^T\,V^\prime \;\; = \;\; -\,Y^T\,B\, Y  \,+\, Y^T\,B\, X_{N_1} \;\;\; .
\end{equation}
But, from eq.~\eqref{eq:vcb}, it follows that $Y^T\,B\,Y \,=\, -\,2\,V_{CB}$ and
eq.~\eqref{eq:vn1} and the fact that the matrix B is symmetric gives $Y^T\,B\, 
X_{N_1}\,=\, -\,2\,V_{N_1}$. Therefore, we reach the conclusion that
\begin{equation}
V_{CB}\;-\;V_{N_1} \;\; = \;\;\frac{1}{2}\,Y^T\,V^\prime \;\; = \;\;
\frac{M^2_{H^\pm}}{2\,v^2} \;\left[ (v^\prime_1\,v_2
\;-\;v^\prime_2\,v_1)^2\; + \; \alpha^2\,v_1^2\right]\;\;\; .
\label{eq:difp}
\end{equation}
Then, it is clear that, if $N_1$ is a minimum of the theory, all of its 
squared masses will be positive, and therefore we will have $V_{CB}\;-\;V_{N_1} 
\;> \;0$, which implies that the charge breaking stationary point, when it 
exists, is always located above the $N_1$ minimum. Furthermore, it is easy to 
obtain the equality $Y\,=\,X\,-\, B^{-1}V^\prime$, so that 
\begin{equation}
Y^T\,V^\prime\;\;=\;\; -{V^\prime}^T\,B^{-1} \,V^\prime\;\;\;.
\label{eq:bm1}
\end{equation} 
In ref.~\cite{pot10} we demonstrated that the matrix $B$ determines the nature 
of the CB stationary point. Notice now that eq.~\eqref{eq:difp} implies that 
$Y^T\,V^\prime\,>\,0$, and from~\eqref{eq:bm1} follows that there is a vector
($V^\prime$, in fact) for which we have ${V^\prime}^T\,B^{-1} \,V^\prime\,<\,0$.
As such, the matrix $B^{-1}$ - and thus, the matrix $B$ itself - is {\em not} 
positive definite. For reasons explained in~\cite{pot10} it cannot be 
negative definite (which arises from requiring that the potential we are working
with is bounded from below), which implies that $B$ is neither positive nor 
negative definite. As a result, the CB stationary point is a saddle point.   

Now we turn our attention to the $N_2$ minimum. {\em A priori} there is no 
reason why the 2HDM potential cannot have, simultaneously, both ``normal" 
minima, so the question arises, can the potential be in an $N_2$ minimum that is
not deeper than a CB stationary point? The answer is no, and the demonstration 
follows very closely the one we just concluded. For the $N_2$ minimum, the 
fields that have non-zero vevs are $\varphi_5 = v^{\prime\prime}_1$, $\varphi_6 
= v^{\prime\prime}_2$ and  $\varphi_7 = \delta$, so that the $X$ vector is now 
given by 
$X_{N_2}\,=\, ({v^{\prime\prime}_1}^2\,+\,\delta^2\,,\,{v^{\prime\prime}_2}^2\,,
\,v^{\prime\prime}_1\, v^{\prime\prime}_2\,,\, -\,v^{\prime\prime}_2\,\delta)$. 
Solving the stationarity conditions as before, we find that the vector 
$V^\prime\,=\, A\,+\,B\, X_{N_2}$, at this minimum, is given by
\begin{equation}
V^\prime_{N_2} \; = \; \begin{bmatrix} V^\prime_1 \\ V^\prime_2 \\ V^\prime_3 \\
V^\prime_4 \end{bmatrix} \; = \; -\frac{\left(V^\prime_3\right)_{N_2}}{2 
v^{\prime\prime}_1 v^{\prime\prime}_2} \, \begin{bmatrix}
x_2 \\ x_1 \\ - 2\, x_3 \\ -2\, x_4 \end{bmatrix} \;\; ,
\end{equation}
and in fact this final expression also applies to the vector $V^\prime$, 
evaluated at the $N_1$ minimum. We still have $X_{N_2}^T\,V^\prime_{N_2}\,=\,0$
and, from eq.~\eqref{eq:mm}, $-\left(V^\prime_3 \right)_{N_2}/(2 
v^{\prime\prime}_1 v^{\prime\prime}_2)\;=\;\left(M^2_{H^\pm}/v^2\right)_{N_2}$.
In this expression the charged scalar mass is the non-zero eigenvalue of the 
matrix~\eqref{eq:mm} at the $N_2$ minimum and $v^2$ is now given by $v^2\,=\,
{v^{\prime\prime}}^2_1\,+\,{v^{\prime\prime}}^2_2\,+\,\delta^2$. We are 
therefore in the exact conditions of the $N_1$ minimum and conclude, likewise, 
that 
\begin{equation}
V_{CB}\,-\,V_{N_2} \; = \;\frac{1}{2}\,Y^T\,V^\prime \; = \;
\left(\frac{M^2_{H^\pm}}{2\,v^2}\right)_{N_2} \left[ (v^\prime_1\,
v^{\prime\prime}_2 \,-\,v^\prime_2\,v^{\prime\prime}_1)^2\, + \, \alpha^2\,
({v^{\prime\prime}_1}^2\,+\,\delta^2)\,+\,\delta^2\,{v^\prime}^2_2\right]\;>
\;0 .
\label{eq:difn2}
\end{equation}
Again, the charge breaking stationary point lies above the normal minimum, and 
again it is a saddle point, for the same reasons we have explained before. 
On the other hand, if neither $N_1$ nor $N_2$ are minima, then it is not 
possible to conclude that $Y^T\,V^\prime\,>\,0$, and then $B$ can be positive
definite and the CB stationary point a minimum. A sufficient condition to 
prevent that, then, is to choose the $b_{ij}$ such that $B$ is {\em not} 
positive definite. 

Unfortunately we cannot apply the procedure followed thus far to determine 
whether one of the minima $N_1$ or $N_2$ is deeper than the other. First, 
though, let us clarify a subtle point: the authors of ref.~\cite{hab} have shown
that within this potential it is always possible to pass from an $N_2$-type 
stationary point to an $N_1$ one, through an appropriate choice of field basis. 
This is accomplished in a simple manner, given that the main difference between
the $N_1$ and $N_2$ stationary points is that, in the latter, the $\Phi_2$  
field has a complex vev. Through a careful choice of basis that complex phase
can be absorbed into the potential's parameters, and we can then proceed to 
minimise the potential, obtaining vevs which are real - meaning, a stationary
point like $N_1$. The key point, though, is that this change of basis does not
affect the value of the potential. Before the change of basis, the potential had
two stationary points, for which the values of the potential were, respectively,
$V_{N_1}$ and $V_{N_2}$. After the change of basis, the potential has new 
parameters, and again, two stationary points, one of the type $N_1$ and another 
of the type $N_2$. What happens is that the value of the new potential at the 
$N_1$ stationary point equals the value of the old potential at $N_2$: 
$V_{N_1}^{new} \,=\,V_{N_2}$, and also, $V_{N_2}^{new} \,=\, V_{N_1}$. The 
possibility of absorbing the complex phase of the $\Phi_2$ vev through a basis 
change and thus passing from an $N_2$ stationary point to a $N_1$ one does not 
mean that $V_{N_1}$ and $V_{N_2}$ are the same. 

If one follows the steps we have outlined for the previous cases, one is left 
with
\begin{equation}
V_{N_2}\;-\;V_{N_1} \;\; = \;\;\frac{1}{2}\left[
\left(\frac{M^2_{H^\pm}}{v^2}\right)_{N_1}\;-\;\left(\frac{M^2_{H^\pm}}{v^2}
\right)_{N_2} \right] \;\left[(v^{\prime\prime}_1\,v_2 \;-\;
v^{\prime\prime}_2\,v_1)^2\; + \; \delta^2\,v_2^2\right]\;\;\; .
\label{eq:difcp}
\end{equation}
The sign of $V_{N_2}\;-\;V_{N_1}$ is thus not defined. If both $N_1$ and $N_2$ 
were minima, the terms proportional to $M^2_{H^\pm}$ would be positive, but it 
is {\em a priori} impossible to tell which one is the largest. We emphasise, 
though, that we in fact do not know, nor have we shown, that such a situation is
indeed possible. It may be that it is in fact impossible to have $M^2_{H^\pm}\,
>\,0$ for both $N_1$ and $N_2$ stationary points simultaneously. Further study 
on this subject is thus necessary. 

For the special case of a potential without explicit CP breaking, the $N_2$
stationary point is the one with spontaneous CP breaking. The $N_1$ stationary
point preserves both charge and CP and it is what we called, in 
ref.~\cite{pot10}, the normal minimum. In that reference we calculated the mass 
matrices for the several minima possible and showed that 
$(M^2_{H^\pm}/v^2)_{N_2}\;=\;-\,b_{44}$.  At the normal minimum we have $M^2_A
\;=\;M^2_{H^\pm}\,+\,b_{44}\,v^2$, $M^2_A$ being the squared mass of the 
pseudoscalar. Then, in this case, eq.~\eqref{eq:difcp} gives the difference of 
the values of the potential at the spontaneous CP breaking stationary point, 
$V_{CP}$, and at the normal one, $V_{N}$, and reduces to
\begin{equation}
V_{CP}\;-\;V_{N} \;\; = \;\;\frac{M^2_A}{2\,v^2}\;\left[
(v^{\prime\prime}_1\,v_2 \;-\;v^{\prime\prime}_2\,v_1)^2\; + \; \delta^2\,v_2^2
\right]\;\;\; .
\label{eq:ma}
\end{equation}
It is clear that if there exists a normal minimum, $M^2_A$ is positive and
the CP stationary point is above the normal minimum. 

It is interesting to point out the following aspect of these results. If one 
observes equations~\eqref{eq:difp} and~\eqref{eq:difn2}, one sees
that the difference in the depth of the potential between the normal minimum and
the CB stationary point is ``controlled" by the charged higgs squared mass. On 
the other hand, equation~\eqref{eq:ma} shows that the potential depth 
difference between the CP and the normal stationary points is proportional to 
the pseudoscalar squared mass. That is, the depth of the potential at a 
stationary point that breaks a given symmetry, relative to the normal minimum,
depends, in a very straightforward way, on the mass of the scalar particle 
directly linked with that symmetry. 

In conclusion we have shown that, for the most general potential of the 2HDM, 
once a minimum that preserves the conservation of electric charge is found, it 
is certainly deeper than any stationary point that breaks charge conservation.
Furthermore, any CB stationary point is, in those circumstances, necessarily a
saddle point. This ensures the stability, against charge breaking, of the 
``normal" minima, at least at tree level. A simple condition to ensure that
charge breaking does not occur, then, is to choose the $b_{ij}$ couplings such
that the $B$ matrix is not positive definite. For the explicitly CP breaking
potential the two possible charge-conserving minima are on equal footing, 
nothing in the model seems to prefer one over the other. We also
showed that the difference in the values of the potential between normal and 
CB stationary points is given, essentially, by the squared charged higgs mass. 
For the special case of a CP conserving potential, that difference between 
the potential at a spontaneous CP breaking stationary point and the normal one
is regulated by the pseudoscalar mass. Thus, the difference in depths of the 
potential between two stationary points with different symmetries is directly
linked to the mass of the scalar particle related to the symmetry being broken.

\vspace{0.25cm}
{\bf Acknowledgments:} We thank Jo\~ao Paulo Silva for many stimulating 
discussions. This work is supported by Funda\c{c}\~ao para a Ci\^encia e 
Tecnologia under contract PDCT/FP/FNU/50155/2003 and POCI/FIS/59741/2004. P.M.F.
is supported by FCT under contract SFRH/BPD/5575/2001.

\end{document}